\documentclass[12pt,preprint]{aastex}

\newcommand\Msun{\mbox{$M_\sun$}}

\newcommand\Lsun{\mbox{$L_\sun$}}

\usepackage{natbib}
\usepackage{amssymb}
\usepackage[figtopcap]{subfigure}
\usepackage{textcomp}

\begin{document}

\shortauthors{Gei{\ss}ler et al.}
\shorttitle{Proper Motion Companion to HII 1348}

\title{A Substellar Common Proper Motion Companion to the Pleiad HII 1348}

\author{Kerstin Gei{\ss}ler, Stanimir A.\ Metchev}
\email{geissler@mail.astro.sunysb.edu}
\affil{Department of Physics and Astronomy, Stony Brook University, 
Stony Brook, NY 11794-3800, USA}
\author{Alfonse Pham}
\affil{Center for Exploration of Energy and Matter, Indiana University, Bloomington, 
IN 47408-1398, USA}
\author{James E.\ Larkin}
\affil{Department of Physics and Astronomy, University of California, Los 
Angeles, California 90095--1562}
\author{Michael McElwain} 
\affil{Astrophysics Science Division, NASA Goddard Space Flight Center, Laboratory for Exoplanets and Stellar Astrophysics, Greenbelt, MD 20771, USA}
\author{Lynne A.\ Hillenbrand}
\affil{Department of Physics, Mathematics \& Astronomy, MC 105--24, 
California Institute of Technology, Pasadena, California 91125}

\begin{abstract}
We announce the identification of a proper motion companion to the star HII~1348, a 
K5\,V member of the Pleiades open cluster.  The existence of a faint point source 
1$\farcs$1 away from HII~1348 was previously known from adaptive optics imaging by 
Bouvier et al.  However, because of a high likelihood of background star 
contamination and in the absence of follow-up astrometry, Bouvier et al.\ tentatively 
concluded that the candidate companion was not physically associated with HII~1348.  
We establish the proper motion association of the pair from adaptive optics imaging 
with the Palomar 5~m telescope.  Adaptive optics spectroscopy with 
the integral field spectrograph OSIRIS on the Keck 10~m telescope reveals that the 
companion has a spectral type of M8$\pm$1.  According to substellar evolution 
models, the M8 spectral type resides within the substellar mass regime at the age of the 
Pleiades.  The primary itself is a known double-lined spectroscopic binary, which makes 
the resolved companion, HII~1348B, the least massive and widest component of this 
hierarchical triple system and the first substellar companion to a stellar primary in the 
Pleiades.

\end{abstract}


\section{INTRODUCTION}

As one of the nearest young \citep[125~Myr;][]{stauffer1998} open clusters, the Pleiades 
have long been recognized as an important astrophysical laboratory for studying stellar 
evolution and the dynamics of stellar associations. Multiplicity studies of the Pleiades 
have focused both on stellar \citep[e.g.,][]{stauffer1984, mermilliod1992, bouvier1997} 
and substellar \citep[e.g.,][]{martin2000, bouy2006} members of the cluster.  However, 
mixed star--brown dwarf systems have not been identified. One of the most extensive 
studies is the adaptive optics imaging survey of \citet{bouvier1997}. Conducted in the 
near-IR, it covered 144 G and K stars to a relatively shallow depth. As a result, its 
sensitivity encompassed only stellar and massive substellar companions. A systematic 
high-contrast imaging survey of the Pleiades on a similar scale but at a higher 
sensitivity has not been performed since. Consequently, no substellar companions are known 
to $>$0.2~\Msun\ stars in the Pleiades. With the frequency of wide substellar companions 
to field-aged Sun-like stars now estimated at $\approx$\,3\,\% 
\citep[0.012\,--\,0.072~\Msun\ brown dwarfs in 28\,--\,1590\,AU orbits;][]{metchev2009}, the 
frequency of brown dwarf secondaries around Sun-like stars in the Pleiades is expected to 
be comparable.

In the present paper, we announce the identification of a low mass companion to the Pleiad 
HII~1348, a K5\,V double-lined spectroscopic binary \citep[hereafter refered to as the 
primary or HII~1348A]{queloz1998}. The faint companion, 
HII~1348B, was already detected by \citet{bouvier1997}.  However, without follow-up 
astrometric observations and due to a non-negligible probability of background star 
contamination, \citet{bouvier1997} conservatively assumed that the candidate companion was 
an unrelated background star.  The astrometric measurements confirm the proper motion 
association of the pair, and AO spectra obtained at Palomar and Keck reveal that the 
companion has a spectral type of M8. 

\section{OBSERVATIONS AND DATA REDUCTION \label{sec_observations}}

\subsection{Astrometry and Photometry}

\begin{figure}
\caption{\label{fig:hii1348_ks_palao}$K_s$-band image of HII~1348A/B taken with the AO 
system on the Palomar 5~m telescope.  The Strehl ratio is $\approx$40\%, and the FWHM of 
the PSF is 0$\farcs$10. The companion is indicated with the white arrow. Total exposure 
time is 21~sec, taken as five 4.2~sec exposures.}
\center
\includegraphics[angle=0,scale=0.4]{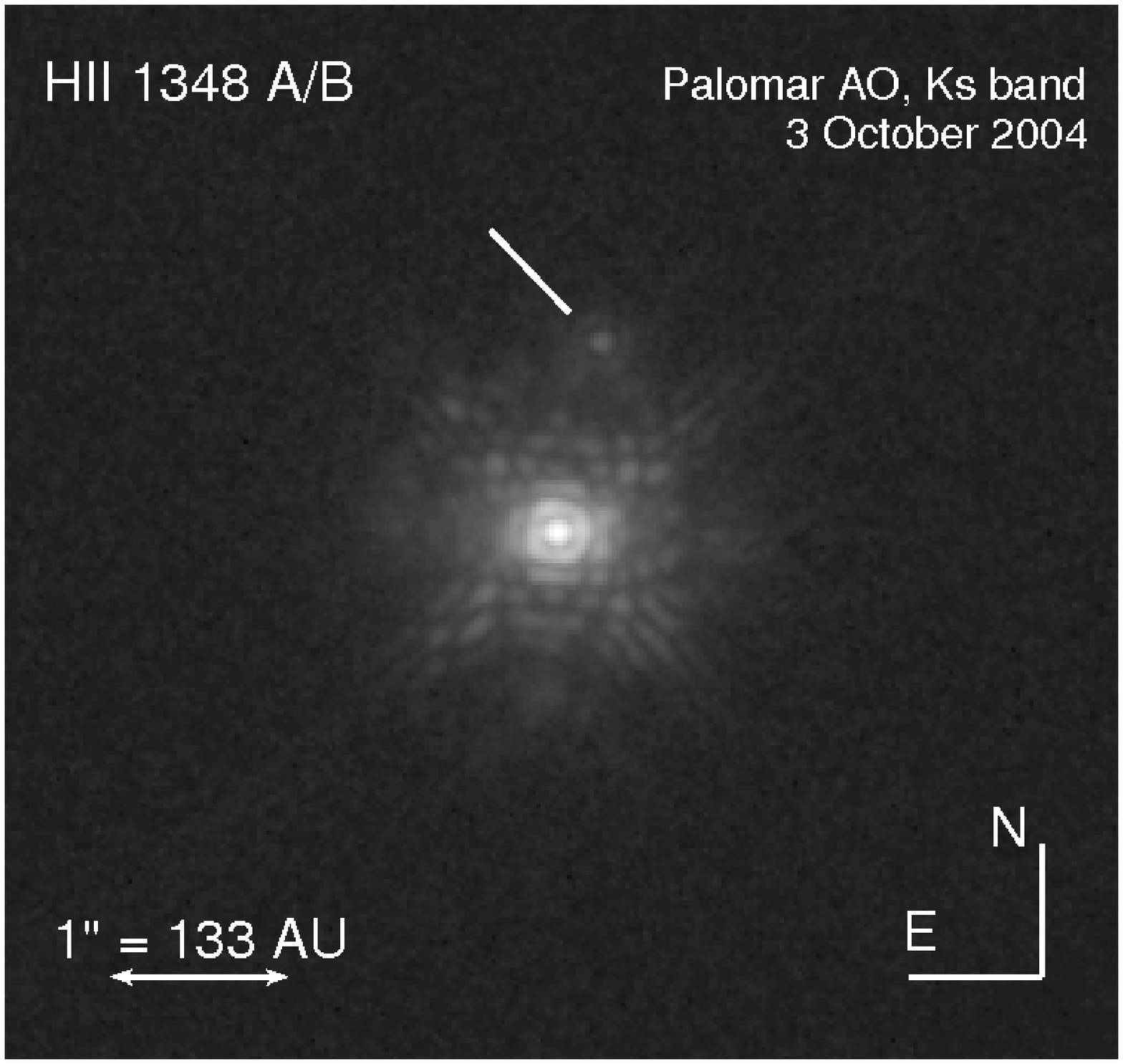}
\end{figure}

HII~1348 ($V=12.6$~mag) was targeted in conjunction with a large-scale natural-guide star 
AO imaging survey of young Sun-like stars, conducted with the PHARO camera 
\citep{hayward2001} on the Palomar Hale telescope.  The data acquisition and reduction 
followed standard near-IR imaging practices and are described in \citep{metchev2004, 
metchev2006}. Diffraction-limited $J$, $H$, and $K_s$ AO imaging of HII~1348 was taken at 
Palomar on 3 October 2004,  in which the tertiary companion is visible $\sim$1$\farcs$1 to 
the North. (Fig.~\ref{fig:hii1348_ks_palao}). To determine the astrometric association of 
the system we used a precise calibration of the pixel scale of the PHARO camera 
\citep{metchev2006, metchev2009}, obtained over time from observations of suggested 
astrometric calibration systems from the Sixth Catalog of Orbits of Visual Binary Stars 
\citep{hartkopf_mason2011}.

HII~1348 was also targeted as part of a demonstration project for the then 
newly-commissioned OSIRIS integral field spectrograph \citep{larkin2006}.  We obtained 
1.18--1.35\,$\micron$ ($Jbb$) and 1.47--1.80\,$\micron$ ($Hbb$) diffraction-limited 
integral field spectra of the pair with the Keck AO system \citep{wizinowich2000} on 21 
November, 2005.  We used the 35\,mas lenslet scale, which allowed a 
0$\farcs$56\,$\times$\,2$\farcs$24 field of view (FOV) with the custom broad band J 
($Jbb$) and H ($Hbb$) filters in OSIRIS.  The spectroscopic reduction of the integral 
field data cube is detailed in \S\ref{sec:ifu}.  Here we only describe the use of the 
OSIRIS $Hbb$\,-band data for astrometry.

\begin{figure}
\caption{\label{fig:hii1348_h_osiris}A two-dimensional rendition of the three-dimensional 
($x, y, \lambda$) Keck/OSIRIS $Hbb$-band data cube, collapsed along the wavelength 
direction.  The 35~mas pixel size under-samples the FWHM=35~mas Keck AO PSF.  The total 
exposure time is 30~min, taken in five exposures of 6~min.}
\center
\includegraphics[angle=0,scale=0.4]{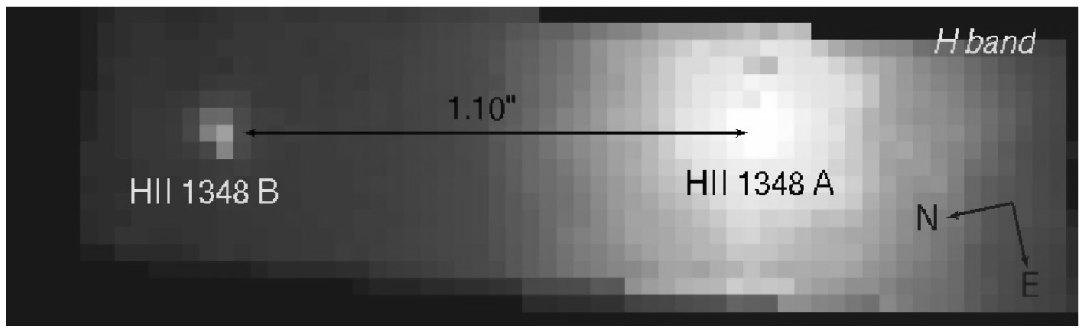}
\end{figure}

The 35\,mas scale of OSIRIS significantly under-samples the 35~mas width of the 
diffraction limited Keck AO PSF in the $Hbb$ filter (Fig. \ref{fig:hii1348_h_osiris}), and 
hence sub-pixel precision astrometry requires sub-pixel dithers.  We did not perform such 
dithers during data acquisition.  However, during the data analysis we noted that the 
spatial positions of the binary components varied monotonically between the short- and the 
long-wavelength end of the $Hbb$ data cube, the difference spanning six lenslets 
(0$\farcs$2) across the 1651 wavelength channels. Since the observations were conducted at 
an airmass of $\sim$1.05 differential atmospheric refraction is negligible and the effect 
is caused by differential refraction arising from the wavefront sensor dichroic in the 
Keck AO system. (The dichroic is oriented at $\approx$45$\degr$ with respect to the 
telescope optical axis, and hence the transmitted science light is refracted in a 
wavelength-dependent fashion.) The benefit to us was the resulting spatial shift  very 
gradually and finely sampled the lenslet size, and hence easily allowed sub-pixel 
precision astrometric measurements.  The lenslet scale and orientation of the OSIRIS 
integral field spectrograph have not been calibrated on sky. We estimated systematic 
uncertainties of 1\% in the lenslet scale and 0.3$\degr$ in the detector orientation. 
These increased the overall positional errors determined from the OSIRIS data by a factor 
of $\sim$50\%.

\begin{figure}
\caption{\label{fig:astrometry}Astrometric measurements of the position of HII~1348B 
relative to its primary star.  The solid points represent measurements at three different 
epochs, spanning ten years.  The dotted curve traces the expected relative motion if the 
companion were an unrelated background star.  The open circles with larger errorbars 
represent the expected location of such a background star at epochs 2 and 3.}
\center
\includegraphics[angle=0,scale=0.4]{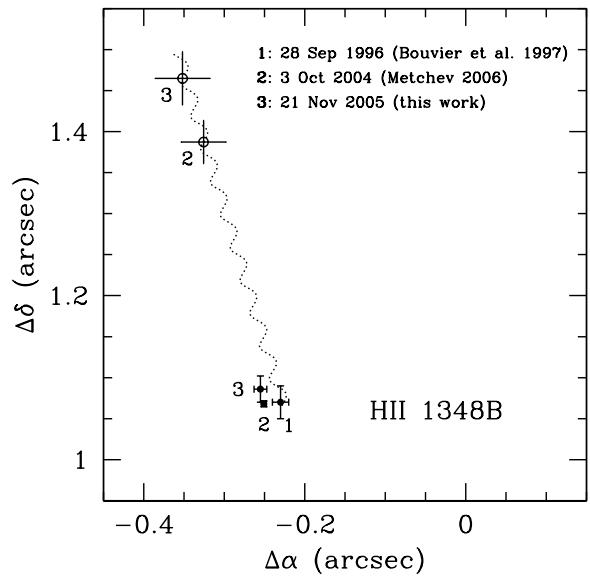}
\end{figure}

Relative photometry and astrometry for the HII~1348A/B pair are given in 
Table~\ref{tab:phot}.  The angular offset and separation between the two components have 
not changed significantly since the discovery of the candidate companion in 1996 
\citep{bouvier1997}, whereas a much more significant change would have been expected if 
the faint candidate companion were an unrelated background star 
(Fig.~\ref{fig:astrometry}).

\subsection{Spectroscopy: Long-slit}

We obtained $K$-band long-slit spectra of HII~1348A and B with the Palomar AO system 
\citep{dekany1998, troy2000} and PHARO \citep{hayward2001} on 3 October 2004. The system 
was aligned along the 0$\farcs$13-wide slit, and was nodded once along an ABC-CBA pattern, 
for an exposure totaling 60~min. After pair-wise subtraction, a first-order polynomial was 
fit to the trace of the primary, which was used to extract the spectra of both visual 
components. The extraction apertures were 0$\farcs$40 and 0$\farcs$12 wide for the primary 
and tertiary, respectively.  The FWHM of the PSF was 0$\farcs$10.  Since HII~1348B lies in 
the halo of the $\sim$5~mag brighter HII~1348Aab, we estimated and subtracted the flux 
from the halo of the primary at the location of the tertiary component by locally fitting 
a quadratic polynomial as a function of both position and wavelength to the radial profile 
of the halo between 0$\farcs$12--0$\farcs$56 from HII~1348B.

\begin{figure}
\caption{\label{fig:hii1348b_long}Long-slit $K$-band spectrum of HII~1348B, smoothed to a 
resolution of R$\sim$250. Comparison spectra of GL\,644C (M7), GL\,752B (M8), LHS\,2924 
(M9), 2MASSJ0746+2000AB (L0.5), and 2MASSJ0208+2542 (L1) are from the IRTF Spectral 
Libary \citep{cushing2005, rayner2009} and have been smoothed to the same resolution.}
\center
\includegraphics[angle=0,scale=0.4]{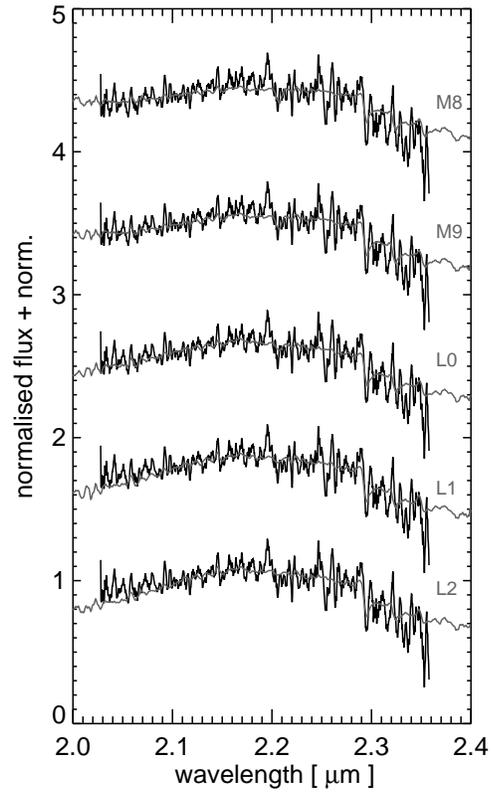}
\end{figure}

The wavelength dispersion for the individual extractions was calibrated using the 
wavelengths of night-sky OH lines.  After wavelength calibration, the individual extracted 
spectra were median-combined, smoothed to the 3.25~pix ($R\approx$1500) resolution of the 
spectroscopic slit, and then the combined spectrum of HII~1348B was divided by the 
combined spectrum of the primary to remove telluric features.  Given the 
relatively low dispersion of our observations, the SB2 nature of the primary was not 
evident in our spectra, and did not affect our telluric calibration.  The systemic 
spectral type of the primary is K5 \citep[$B-V=1.18$~mag;][]{johnson1958, herbig1962}, and 
so the corrected spectrum of the tertiary component was multiplied by a 
synthetic K5\,V stellar spectrum.  The final reduced $K$-band spectrum of HII~1348B is 
shown in Figure~\ref{fig:hii1348b_long}. The long-slit K band spectrum closely resembles 
the SEDs of late-M and early-L type dwarfs. Using a $\chi^2$ minimization approach we fit 
the K band spectrum to the standards, limiting the fit to the 2.05-2.30\,$\mu$m wavelength 
region. The long-slit K band spectrum is best-fit by the L1 spectral standard. 
Nevertheless, since the M9 standard returns almost as good a fit as the L1 standard, we 
adopt a median spectral type of L0$\pm$1. This marginally disagrees with the IFU spectra 
presented next, that indicate a slightly earlier, M8$\pm$1, spectral type. The discrepancy 
is discussed in \S~\ref{sec:spt}.

\subsection{Spectroscopy: Integral Field}
\label{sec:ifu}

The Keck AO/OSIRIS integral field spectra of HII~1348A/B were obtained with the system 
aligned along the long side of the $0\farcs56\times2\farcs24$ FOV.  We obtained five 
300~sec exposures at $Hbb$ and four 300~sec exposures at $Jbb$, with an extra 300~sec 
exposure on sky at each band for sky subtraction.  An A0 star, HD~24899, was observed 
immediately afterwards in each band for telluric calibration.

The data were reduced with the OSIRIS data reduction pipeline 
(DRP)\footnote{\url http://www2.keck.hawaii.edu/inst/osiris/tools/.}, following the 
procedure described in \citet{mcelwain2007}.  We used the default 7~pix (0$\farcs$245) 
radius aperture to extract the spectra of HII~1348A and B, where the center of the 
aperture was varied with wavelength to account for differential refraction arising both 
from the Earth's atmosphere and from the dichroic in the Keck AO system. 

\begin{figure}
\caption{\label{fig:hii1348b_ifu} $Jbb$- and $Hbb$-band IFU spectra of HII~1348B, smoothed 
to a resolution of $R\sim925$. Comparison dwarfs are GL\,406 (M6), GL\,644C (M7), GL\,752B 
(M8), LHS\,2924 (M9), and 2MASSJ0746+2000AB (L0.5) from the IRTF Spectral Libary 
\citep{cushing2005, rayner2009}, shown here at a resolution $R=1000$. }
\center
\includegraphics[angle=0,scale=0.4]{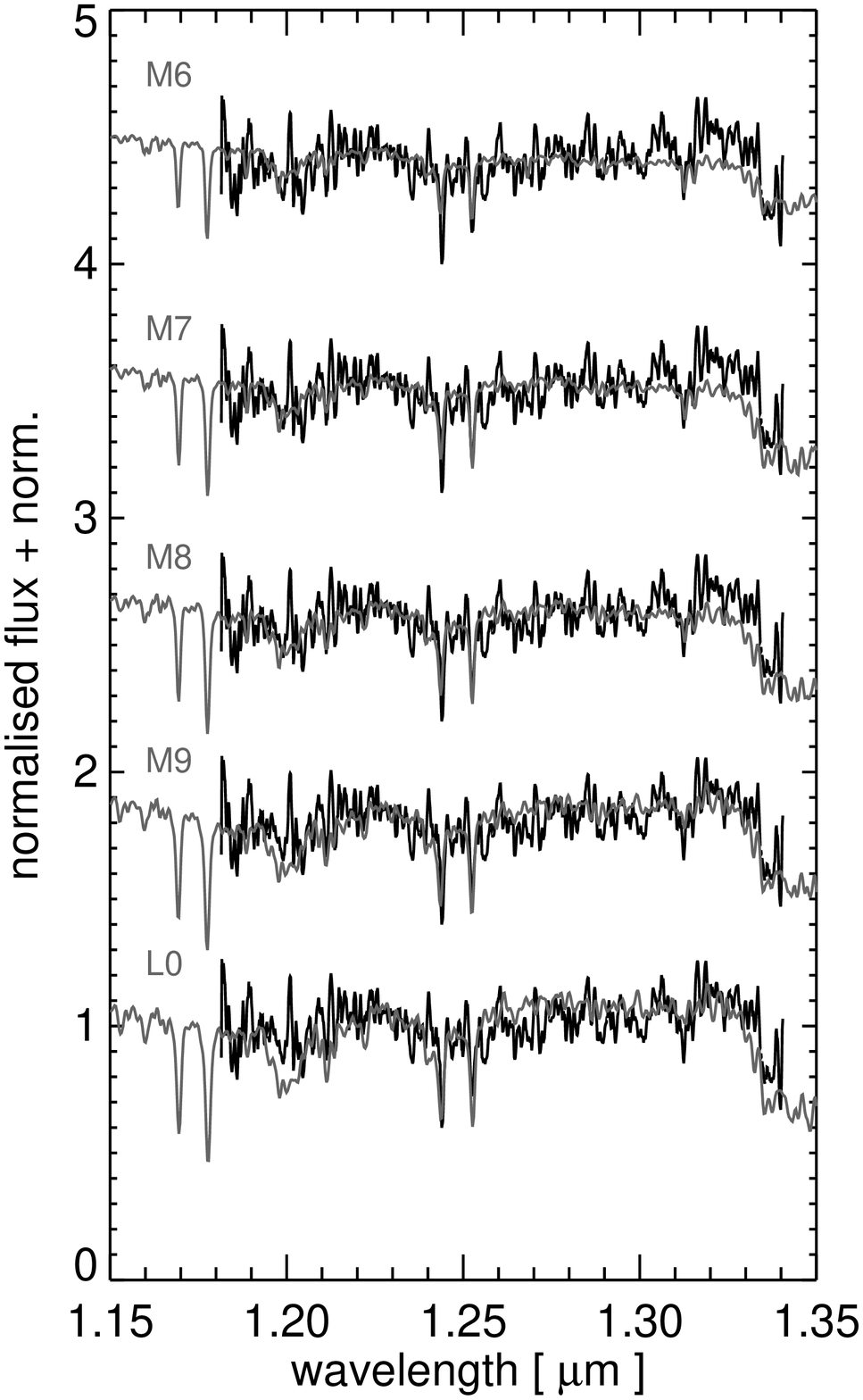}
\includegraphics[angle=0,scale=0.4]{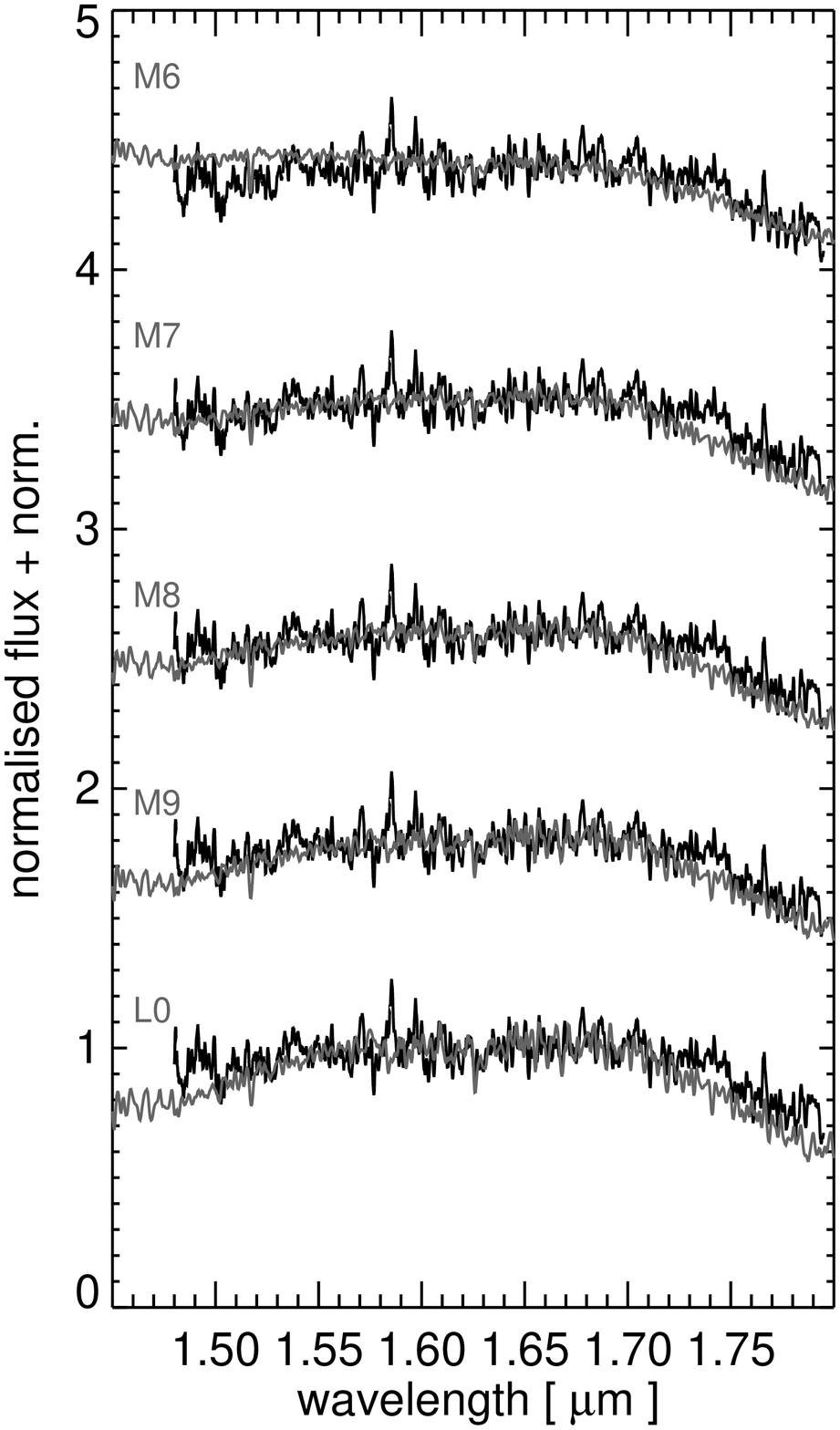}
\end{figure}

Based on the input set of individual exposures, the OSIRIS DRP produces a single 
median-combined, wavelength-calibrated spectrum for each extracted object.  A 
wavelength-collapsed $Hbb$ OSIRIS data cube of HII~1348A/B is shown in 
Figure~\ref{fig:hii1348_h_osiris}. The final $Jbb$ and $Hbb$ spectra of HII~1348B are 
shown in Figure~\ref{fig:hii1348b_ifu}. Comparison spectra of M and L dwarfs are from the 
IRTF Spectral 
Libary\footnote{\url http://irtfweb.ifa.hawaii.edu/\~{}spex/IRTF\_Spectral\_Library/} 
\citep{cushing2005, rayner2009}.

To assess the best-fit spectral types of the $Jbb$ and $Hbb$ spectra, we followed a 
$\chi^2$ minimization when comparing the target spectra and the comparison standards. The 
$\chi^2$ fitting was limited to the 1.185-1.340\,$\mu$m and 1.490--1.800\,$\mu$m 
wavelength regions, respectively, to avoid regions of high noise. The spectral fitting 
yields best-fit spectral types of M9 and M7/M8 for the $Jbb$ and $Hbb$ spectra, 
respectively.

\section{The HII~1348 system}

\subsection{Probability of Physical Association}

While HII~1348A and HII~1348B likely constitute a physically bound system, there is still 
a possibility that HII~1348B may be an unrelated Pleiades brown dwarf. We estimate the 
probability for such a spurious alignment by considering the surface density of Pleiades 
brown dwarfs and the total sky area surveyed by \citet{bouvier1997} in their AO survey of 
the Pleiades. (HII~1348B was the only potential substellar companion found by that survey, 
and we targeted it for confirmation precisely because of its unresolved association 
status.)  

The most comprehensive survey for Pleiades brown dwarfs is that of \citet{moraux2003}, who 
imaged a total area of 6.4 square degrees and discovered 40 brown dwarf (BD) candidates as 
faint as $I\,=\,21.7$~mag, or as low as $\sim$\,0.03\Msun\ in mass \citep{baraffe1998}. 
The total area surveyed by the \citet{bouvier1997} AO observations of 144 Pleiades stars 
is $2.5\times10^{-3}$ square degrees: 144 stars $\times$ a $15\arcsec\times15\arcsec$ area 
imaged around each star. The depth of the \citet{bouvier1997} survey is $K=17$~mag, also 
corresponding to a $\approx$\,0.03\,\Msun\ lower mass limit.  The expected number of 
$\gtrsim$\,0.03\,\Msun\ brown dwarfs in the \citet{bouvier1997} AO survey is then 
$2.5\times10^{-3}$\,sq.\,deg.\,/\,6.4\,sq.\,deg.\,$\times$\,40\,BD's\,=\,0.016\,BD's.  The 
Poisson probability of detecting at least one Pleiades brown dwarf that is not orbiting 
another Pleiades member is then 
$P(n_{\rm BD}\geq1)=1-P(n_{\rm BD}=0)=1-\exp(-0.016)=1.5\%$.

We therefore conclude that the HII~1348 spectroscopic binary and its visual tertiary 
companion form a common proper motion pair, and that there is a 98\% probability that they 
form a physically bound multiple system.  We will henceforth assume that the visual 
companion HII~1348B is indeed bound to HII~1348Aab.

\subsection{Analysis of the spectra}
\label{sec:spt}
Despite the long-slit $K$-band spectrum pointing towards a later spectral type, we adopt a 
spectral type of M8$\pm$1 for HII~1348B: the mean of the OSIRIS $Jbb$- and $Hbb$-band 
spectra.  We note that the $0\farcs$13 slit width for the Palomar AO $K$-band long-slit 
spectrum was only slightly wider than the $0\farcs10$ PSF, and only $\sim$5 times wider 
than the alignment precision for the target on the slit.  Rather than oriented along the 
parallactic angle, the long slit was aligned along the visual binary components. Both 
factors may lead to wavelength-dependent slit looses, either through inadequate centering 
of the target on the slit, or through differential atmospheric refraction.  These incur 
continuum gradients that systematically affect the inferred spectral type of the 
companion, especially in cases where the spectral type is based on the continuum slope or 
on broad band molecular absorption features at either end of the spectrum 
\citep[see discussion in][]{goto2003, mcelwain2007}. A variation in continuum slope could 
also be caused by an increased error in the telluric correction. The thicker atmospheric 
layer at Palomar Observatory causes greater uncertainty in the telluric correction of the 
K band data, compared to the OSIRIS $Jbb$- and $Hbb$-band taken at Mauna Kea, and 
therefore more uncertainty in the spectral type, which is partly based on H$_2$O 
depression.

\begin{figure}
\caption{\label{fig:hii1348b_KI} Zoom of the K~I doublet of HII~1348B, showing an apparent 
redshift in the 1.243$\mu$m line of the doublet, possibly due to a detector cross-talk 
effect.}
\center
\includegraphics[angle=90,scale=0.4]{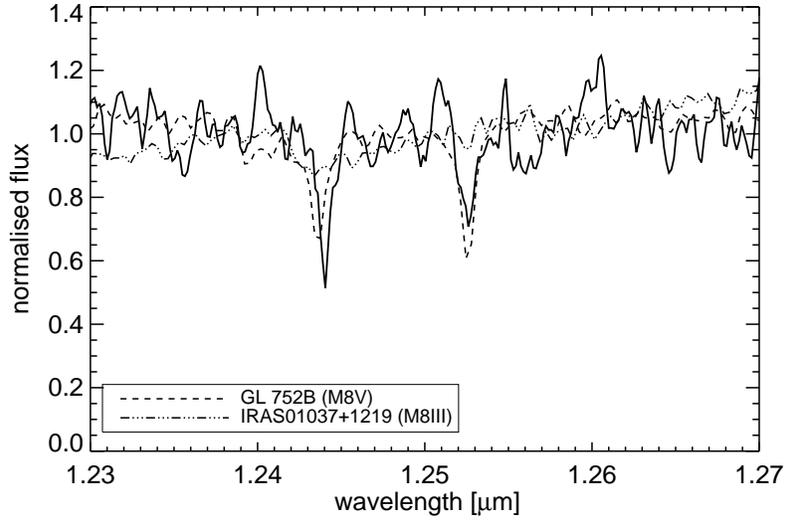}

\end{figure}

The OSIRIS $Jbb$-band spectrum covers the gravity sensitive K~I doublet at 
1.243/1.252\,$\mu$m. In young dwarfs, low gravity causes the alkali lines to appear weaker 
and sharper than in normal (older) dwarfs \citep{steele1995, martin1996, kirkpatrick2008}. 
Thus, by comparing the shape and strengths of the K~I doublet to a normal M8 dwarf, we 
should be able to infer if HII~1348B has lower gravity than the comparison dwarf. 
Figure\,\ref{fig:hii1348b_KI} shows such a comparison. The K~I line at 1.243\,$\mu$m is 
slightly red-shifted and much stronger than expected. The offset in position remains 
unexplained at present, but may have to do with varying contamination from the much 
brighter primary component because of cross-talk in the data from the early commissioning 
days of OSIRIS. Disregarding the K~I line at 1.243\,$\mu$m, the K~I line at 1.252\,$\mu$m 
appears to be intermediate in strength between the giant and the dwarf comparison spectra. 
However, given the fact that the K~I line strength is consistent within the noise of our 
data, we can draw no reliable conclusions on the gravity of HII~1348B.

\subsection{The mass of HII~1348B}

Given the uncertainty in the absolute distance to the Pleiades, with measurements varying 
between 120\,pc and 140\,pc, we consider two distances, 120.2\,$\pm$\,1.9\,pc \citep[based 
on the revised HIPPARCOS measurements by][]{vanleeuwen2009} and 133\,pc \citep [a weighted 
mean of trigonometric and orbital parallax distances from][]{pan2004, munari2004, 
zwahlen2004, southworth2005} when calculating the absolute magnitudes and luminosity of 
HII~1348B (Table \ref{tab:info}).  Estimates for the age of the Pleiades range between 
100~Myr and 125~Myr \citep{meynet1993, stauffer1998}.

The models of \citet{baraffe1998} do not include absolute magnitudes as faint as those of 
HII~1348B at an age of 0.1\,Gyr, implying that the objects mass is below the hydrogen 
burning limit. Averaging the mass estimates \citep{chabrier2000} given by the absolute J, 
H, and K$_S$ magnitudes reported in this paper, yields masses of 0.056$\pm$0.003, and 
0.063$\pm$0.004 at 120\,pc and 133\,pc, respectively. The corresponding luminosity 
estimates from the Lyon models are listed in Table \ref{tab:info}.

\begin{figure}
\caption{\label{fig:hii1348b_evo} Plot of the \cite{burrows1997} and \cite{chabrier2000} 
evolutionary models. The box indicates the allowed range of parameters (age and L$_{bol}$) 
for HII~1348B at a distance of 120\,pc.}
\center
\includegraphics[angle=90,scale=0.29]{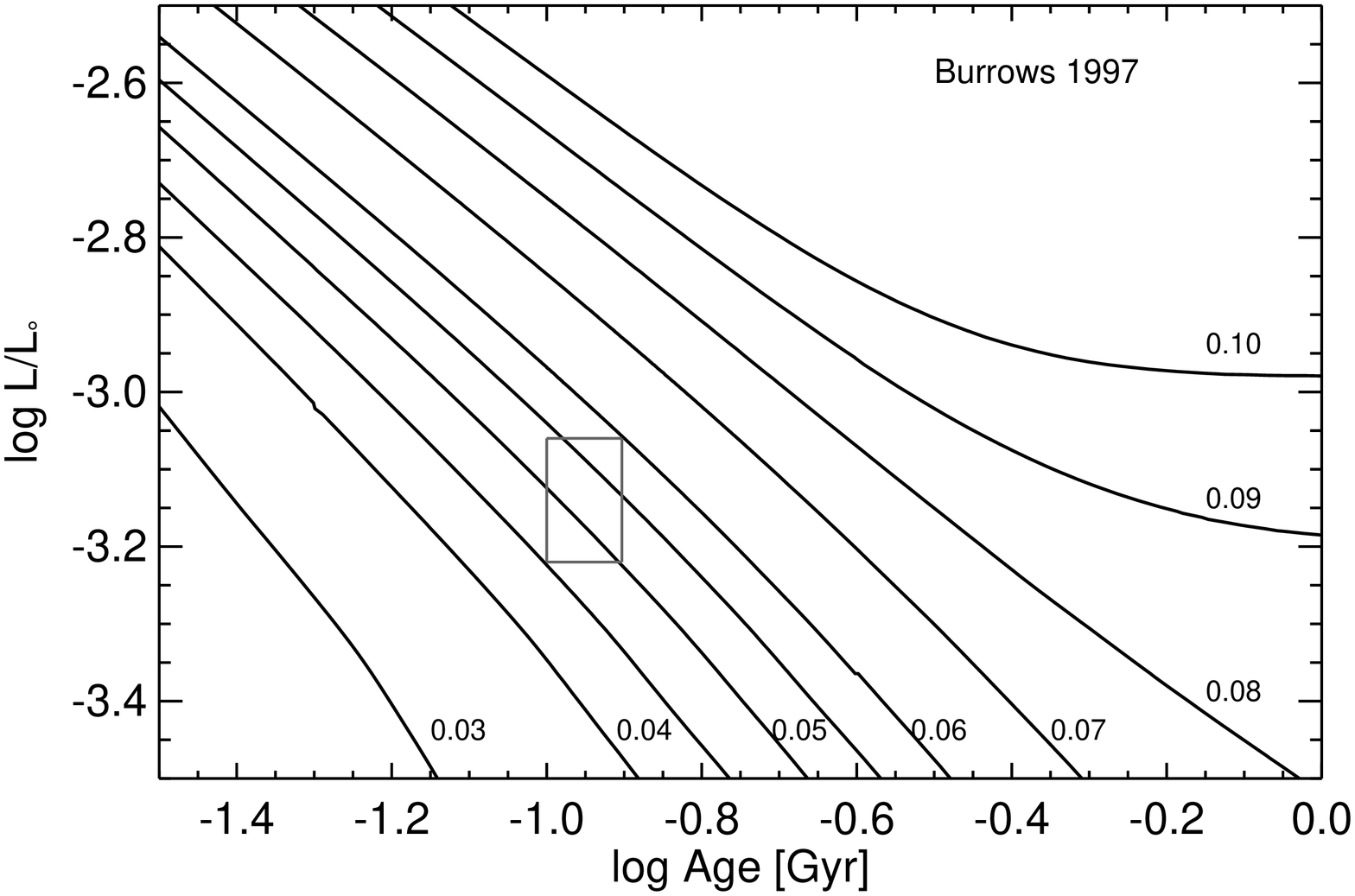}
\includegraphics[angle=90,scale=0.29]{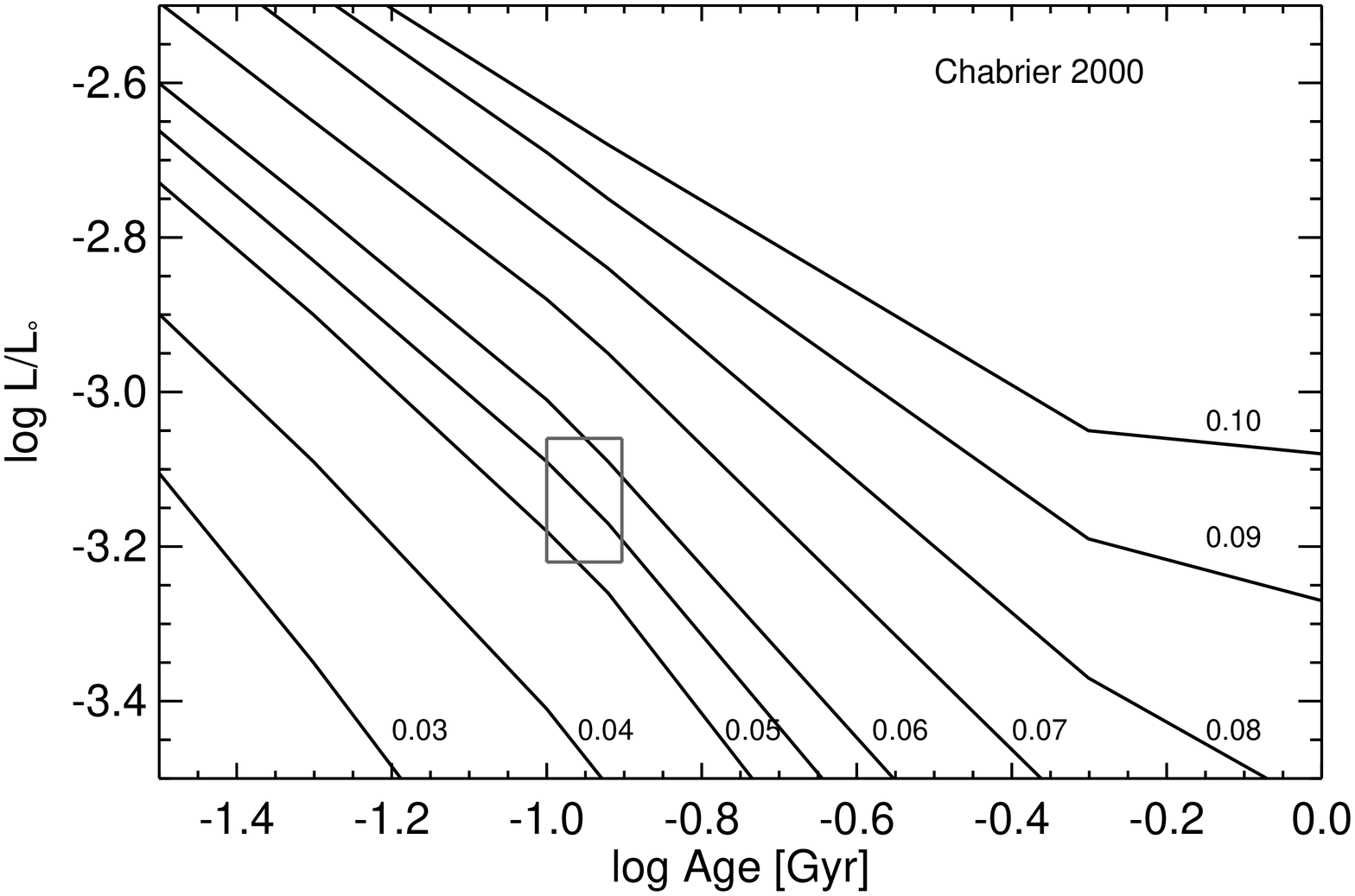}
\end{figure}

Likewise, we can estimate the mass of HII~1348B via its luminosity. Using measured
bolometric corrections for M8-type dwarfs \citep{dahn2002, golimowski2004, leggett2002}, 
we calculated the bolometric magnitude and luminosity of HII~1348B (Table \ref{tab:info}).
At an age of 100--125\,Myr and a distance of 120\,pc, models from \citet{burrows1997} and 
\citet{chabrier2000} yield mass estimates between 0.053--0.055\,$\Msun$ 
(Fig.\,\ref{fig:hii1348b_evo}). 

The mass of HII~1348B is within the range of mass estimates for other Pleiads with similar 
spectral types. \cite{bouvier1998} gives a mass of 0.061\,$\Msun$ (at a distance of 
125\,pc) for the M6 dwarf CFHT~PL~14 \citep{stauffer1998}. PPl~1 is a M6.5$\pm$0.5 dwarf 
with an estimated mass of $\sim$0.074\,$\Msun$ \citep{bihain2010}, and is right at the 
lithium depletion edge in the Pleaides \citep{stauffer1998}. Calar~3 and Teide~1, two M8 
dwarfs, have estimated masses of $\sim$0.054\,$\Msun$ and $\sim$0.052\,$\Msun$ \citep[at 
d=120\,pc;][]{bihain2010}, respectively, and their substellar nature is confirmed through 
the presence of lithium \citep{rebolo1996}.  HII~1348B has the same spectral type as 
Calar~3 and Teide~1, and we therefore conclude that is also substellar.  Future optical 
spectroscopy could confirm the presence of lithium in HII~1348B.

\subsection{Mass ratio}

HII~1348A is a double-lined spectroscopic binary \citep{queloz1998}, and hence an upper 
mass limit for the primary can be obtained by assuming that both components are K5\,V 
stars. With the mass of a single K5\,V star being $\sim$0.65$\Msun$ \citep{zakhozhaj1998}, 
the SB2 star HII~1348Aab has a mass of $\sim$\,1.3$\Msun$. A more precise mass 
estimate can be obtained using the $B-V$ colors of HII 1348Aa and HII 1348Ab 
(1.05 and 1.35, respectively) given by \citet{queloz1998}. The colors roughly 
translate to masses of 0.67$\pm$0.07$\Msun$ and 0.55$\pm$0.05$\Msun$, respectively, 
yielding a total estimated mass of 1.22$\pm$0.09$\Msun$ for HII~1348Aab.  Adopting the 
latter mass for the SB2 component, the mass ratio of HII~1347B to HII~1347Aab is between 
0.043--0.052 (Table \ref{tab:info}).  This is the lowest among the known Pleiad multiples 
\citep{bouvier1997, bouy2006}, and is comparable to that of very low-mass ratio binaries 
in the field \citep{faherty2011}.

\section{DISCUSSION AND CONCLUSION}

HII~1348B is a new M8 brown dwarf member of the Pleiades, and the first substellar 
companion discovered around a Pleiades star.  Given that no other substellar companions 
were discovered in the \citet{bouvier1997} survey at similar or wider separations, it is 
worth considering whether HII~1348B may be unusually weakly bound, compared to other 
binary systems in the Pleiades or in the field.  

\begin{figure}
\caption{\label{fig:Ebind} Comparison of the binding energy of stellar and VLM binary and 
multiple systems \citep{close2003, close2007}. The symbols are: filled triangles - stellar 
binaries and multiples \citep{close1990}, filled circles - low-mass Hyades binaries 
\citep{reid1997}, open stars - VLM binares \citep[][and references therein]{close2003}, 
asterisks - stellar Pleiades binaries \citep{bouvier1997}, open triangles - VLM Pleiades 
binaries \citep{bouy2006}, and open square - HII~1348.}
\center
\includegraphics[angle=90,scale=0.55]{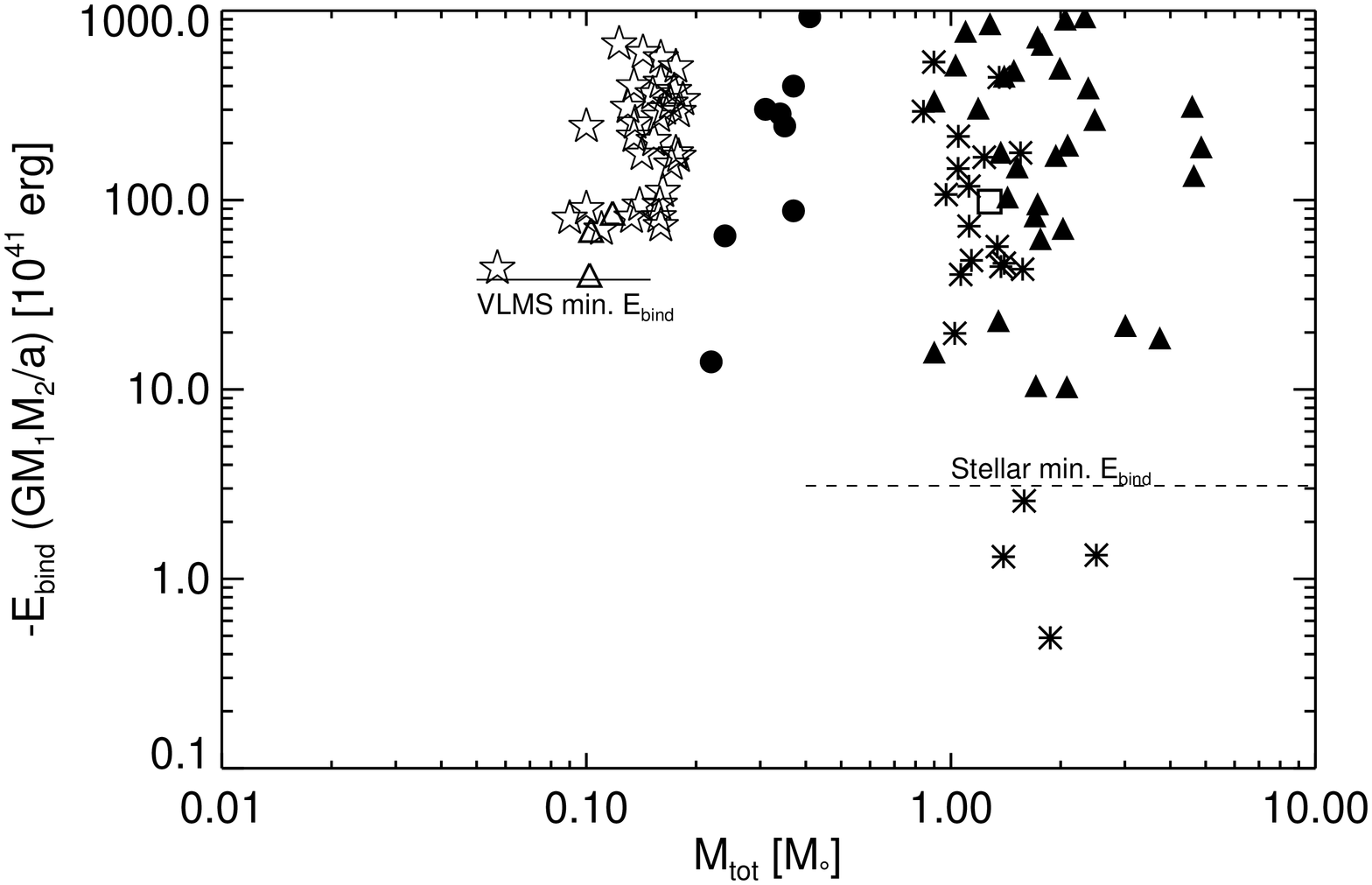}

\end{figure}

\citet{bouvier1997} found a total of 28 stellar binaries in the Pleiades in their CHFT AO 
survey.  HST surveys of very low mass stars and brown dwarfs conducted by 
\citet{martin2003} and \citet{bouy2006} revealed three additional binaries. In Figure 
\ref{fig:Ebind} we compare the binding energies of these systems, and those of field 
A--M binaries \citep{close1990, close2003, close2007}, to the binding energy of 
HII~1348A/B.  As can be seen, HII~1348A/B sits in the middle of the locus for stellar 
binaries, and is comparably or more tightly bound even than the three very low mass 
Pleiades binaries.  What is more, substellar companions up to 10 times further away from 
their primaries would still be well above the minimum stellar binding energy in the 
Pleiades.

The dearth of known brown dwarf companions to stars in the Pleiades may thus be 
attributable to the lack of a follow-up sensitive and comprehensive high-contrast imaging 
survey of the cluster.  Small samples of Pleiades stars have since been observed in deep 
surveys by \citet[23 stars;][]{metchev2009} and \citet[14 stars;][]{tanner2007}, with no 
new brown dwarf companion detections.  However, a much more comprehensive survey is needed 
to reveal brown dwarf companions with any statistical confidence \citep{metchev2009}.  

Current AO systems at Keck or Gemini North should allow the detection of companions with 
masses down to $\sim$0.03\,$\Msun$ at separations larger than $\sim$60AU 
\citep[0$\farcs$5;][]{lafreniere2007, metchev2009b}. Using the Gemini North AO system 
Altair in combination with NIRI, \citet{lafreniere2007} showed that companions up to 
9.5\,mag fainter can be detected at separations of $>$0$\farcs$5 from the primary. 
Accounting for the narrow band filter used during the observation, brown dwarfs down to 
0.03\,$\Msun$ should be detectable around bright (V$<$12\,mag) Pleiades stars. Likewise, 
using angular differential imaging \citep[ADI,][]{marois2006} in combination with the Keck 
AO system, \citet{metchev2009b} demonstrated that companions up to $\Delta$H=10.5\,mag 
fainter than the primary can be detected at separations larger than 0$\farcs$5. Thus, at 
the distance of the Pleiades brown dwarfs down to 0.02\,$\Msun$ should be detectable at 
separations larger than $\sim$60AU. Upcoming instruments like the Gemini planet imager 
\citep{macintosh2008, mcbride2011} and the extreme AO system at Palomar 
\citep[PALM3000,][]{bouchez2008} will enable detections of brown dwarf companions at even 
smaller separation down to 0$\farcs$2.

Given the importance of the Pleiades as an age and a dynamical benchmark for stellar and 
substellar evolution, a deep AO survey could have a high impact in delivering substellar 
objects much fainter and cooler than the ones presently known from wide-area surveys. 
While wide-area surveys are more efficient in discovering large numbers of substellar 
candidate members, the limited precision of seeing-limited astrometry, and the 
increasingly challenging radial velocity measurements at fainter magnitudes, prevent the 
unequivocal confirmation of the lower mass candidates as bona-fide members.  The factor of 
$\sim$100 higher precision attainable over narrow angles with AO leaves little doubt about 
the astrometric association of candidate binaries, and is thus also an excellent tool for 
confirming the cluster membership of even the lowest-mass, faintest companions.

\acknowledgments

{\bf Acknowledgments.}   Partial support for S.A.M.\ was provided through the 
{\it Spitzer} Fellowship Program under award 1273192.  The authors also wish to extend 
special thanks to those of Hawaiian ancestry on whose sacred mountain of Mauna Kea we are 
privileged to be guests.  

\facility{{\it Facilities:} Keck II Telescope, Palomar Observatory's 5~meter Telescope} 


\bibliographystyle{apj}
\bibliography{ms}

\clearpage

\clearpage

\begin{deluxetable}{lrrrrrr}\scriptsize
\tablewidth{0pt}
\tablecaption{Photometry and Astrometry of HII~1348B \label{tab:phot}}
\tablehead{ \colhead{date} & \colhead{$\Delta$K$_S$} & 
	\colhead{K$_S$} & \colhead{J} &
	\colhead{H} & \colhead{$\Delta\alpha$} & \colhead{$\Delta\delta$}  \\
	\colhead{ddmmyy} & \colhead{{[}mag{]}} & \colhead{{[}mag{]}} & 
	\colhead{{[}mag{]}} & \colhead{{[}mag{]}} &
	\colhead{{[} $\arcsec$ {]}} & \colhead{{[} $\arcsec$ {]}} }
\startdata
??/09/96\tablenotemark{\dag} & 5.47 [0.01]\tablenotemark{\ddagger} & 15.06 [0.05] & \nodata & \nodata & $-0.23$\phantom{0}  [0.01]\phantom{0} & 1.07\phantom{0}  [0.02]\phantom{0} \\ 
03/10/04 & 5.15 [0.09] & 14.88 [0.09] & 16.04 [0.09] & 15.30 [0.09] & $-0.251$ [0.003] & 1.068 [0.004] \\
21/11/05 & \nodata & \nodata & \nodata & \nodata & $-0.255$ [0.008] & 1.086 [0.016] \\
\enddata
\tablenotetext{\dag}{The first-epoch data are from \citet{bouvier1997}.  The exact date of 
the observation is not listed, but the data are obtained between 25 Sep 1996 and 1 Oct 
1996.  The R.A.\ and DEC offsets are calculated from the radial separation ($1\farcs09$) 
and position angle (347.9$\degr$) of the companion, assuming uncertainties of 1\% in pixel 
scale and 0.3$\degr$ in orientation.  \citet{bouvier1997} do not list astrometric 
uncertainties.\\
$\ddagger$ \citet{bouvier1997} observations were done in $K$ rather than in
$K_s$ with an apparent $K$-band magnitude of 9.59\,mag for HII 1348A, compared to the 
2MASS $K_s$-band magnitude of 9.719\,mag used in this paper.
Although \citet{bouvier1997} assign an uncertaintity of 0.01 magnitudes to their 
photometry, this is inconsistent with normal accuracy of even relative photometry in 
low-Strehl ratio AO images, and  inconsistent with our higher-quality PALAO images, so we
believe the error is underestimated.}
\end{deluxetable}

\begin{deluxetable}{lrr}
\tablewidth{0pt}
\tablecaption{Parameters of HII~1348B. \label{tab:info}}
\tablehead{\colhead{ } & \colhead{120pc} & 
	\colhead{133pc}}
\startdata	
$m-M$                  & 5.40 [0.03]   & 5.62  [0.03]  \\
$M_J$   {[}mag{]}      & 10.64 [0.10]  & 10.42 [0.10]  \\
$M_H$   {[}mag{]}      &  9.90 [0.10]  &  9.68 [0.10]  \\
$M_K$   {[}mag{] }     &  9.48 [0.10]  &  9.26 [0.10]  \\
$a$ {[}AU{]}           & 131.0 [2.2]   & 145.0 [2.3]   \\
\hline 
\multicolumn{3}{c}{{\it via absolute magnitudes}} \\
log $L/\Lsun$          & -3.16 [0.05]  & -3.05 [0.06]  \\
$M_B$ {[}$\Msun${]}    & 0.056 [0.003] & 0.063 [0.004] \\
$q$		       & 0.046 [0.004] & 0.052 [0.005] \\ 
EB {[}10$^{41}$ erg{]} & 92.0  [8.5]   & 93.5  [9.2]  \\	
\hline
\multicolumn{3}{c}{{\it via bolometric corrections}} \\
log $L/\Lsun$          & -3.14 [0.08]  & -3.06 [0.08]  \\
\multicolumn{3}{c}{\cite{chabrier2000}:} \\
$M_B$ {[}$\Msun${]}    & 0.055 [0.007] & 0.061 [0.008] \\
$q$		       & 0.045 [0.007] & 0.050 [0.008] \\
EB {[}10$^{41}$ erg{]} & 90.4  [13.4]  & 90.6  [13.7]  \\
\multicolumn{3}{c}{\cite{burrows1997}:} \\
$M_B$ {[}$\Msun${]}    & 0.053 [0.007] & 0.057 [0.008] \\
$q$		       & 0.043 [0.007] & 0.047 [0.007] \\ 
EB {[}10$^{41}$ erg{]} & 87.1  [13.3]  & 84.6  [13.5]  \\
\enddata
\end{deluxetable}

\end{document}